\documentclass[titlepage,12pt]{utarticle}
\usepackage{amsmath,amssymb,amsthm,amscd,cite}
\usepackage{latexsym,amsmath,amssymb,epsfig,graphicx}
\usepackage[all]{xy}
\usepackage{epsf}
\xyoption{v2}
\xyoption{2cell}
\xyoption{dvips}
\usepackage{psfrag}
\CompileMatrices
\LaTeXdiagrams
\UseAllTwocells
\newdimen\tableauside\tableauside=1.0ex
\newdimen\tableaurule\tableaurule=0.4pt
\newdimen\tableaustep
\def\phantomhrule#1{\hbox{\vbox to0pt{\hrule height\tableaurule width#1\vss}}}
\def\phantomvrule#1{\vbox{\hbox to0pt{\vrule width\tableaurule height#1\hss}}}
\def\sqr{\vbox{%
  \phantomhrule\tableaustep
  \hbox{\phantomvrule\tableaustep\kern\tableaustep\phantomvrule\tableaustep}%
  \hbox{\vbox{\phantomhrule\tableauside}\kern-\tableaurule}}}
\def\squares#1{\hbox{\count0=#1\noindent\loop\sqr
  \advance\count0 by-1 \ifnum\count0>0\repeat}}
\def\tableau#1{\vcenter{\offinterlineskip
  \tableaustep=\tableauside\advance\tableaustep by-\tableaurule
  \kern\normallineskip\hbox
    {\kern\normallineskip\vbox
      {\gettableau#1 0 }%
     \kern\normallineskip\kern\tableaurule}%
  \kern\normallineskip\kern\tableaurule}}
\def\gettableau#1 {\ifnum#1=0\let\next=\null\else
  \squares{#1}\let\next=\gettableau\fi\next}
\usepackage{color}
\numberwithin{equation}{section}

\def\blue{\color{blue}}
\def\red{\color{red}}
\def\black{\color{black}}

\newcommand{\be}{\begin{equation}}
\newcommand{\ee}{\end{equation}}

\newcommand{\IP}{\mathbb{P}}

\newcommand\IZ{\mathbb {Z}}

\newcommand{\ba}{\begin{array}}
\newcommand{\ea}{\end{array}}

\newcommand{\om}{\overline{M}}

\newcommand{\IF}{{\mathbb F}}

\begin{document}
\preprint{
    {\tt hep-th/0507240}\\
}
\title{
The Ruled Vertex and Nontoric del Pezzo Surfaces}
\author{Duiliu-Emanuel Diaconescu\footnote{{\tt duiliu@physics.rutgers.edu}}~~and 
Bogdan Florea\footnote{{\tt florea@physics.rutgers.edu}}}
\oneaddress{
      \smallskip
      {\centerline {\it  Department of Physics and Astronomy, 
Rutgers University,}}
      \smallskip
      {\centerline {\it Piscataway, NJ 08854-8019, USA}}}

\date{July 2005}

\Abstract{
We construct the topological partition function of local nontoric del Pezzo
surfaces using the ruled vertex formalism.  
}
\maketitle 

\section{Introduction}

Let $S_k$ be the del Pezzo surface obtained by blowing up $k$ generic points on $\IF_0$, 
$3\leq k \leq 7$, and let $X_k$ be the total space of the canonical bundle $K_{S_k}$; 
$X_k$ is a noncompact Calabi-Yau threefold. 
In this note we compute the partition function of the topological {\bf A}-model 
with target space $X$ using complex structure deformations and 
the ruled vertex formalism developed in \cite{DFS}. 

The topological {\bf A}-model partition function or, equivalently, the Gromov-Witten theory 
of $X_k$ is defined by summing over stable maps to $X_k$. Since $S_k$ is Fano, any such map 
must factor through the zero-section of the fibration $X_k \to S_k$, therefore the Gromov-Witten 
theory of $X_k$ is well defined. We have 
\be\label{eq:GWA}
Z_{X_k} = \sum_{g\in \IZ}\sum_{\stackrel{\beta\in H_2(S_k)}{\beta\neq 0}} 
g_s^{2g-2} q_{S_k}^{\beta}\int_{[\om_{g,0}(S_k,\beta)]^{vir}} 
e(R^1\rho_*ev^*K_{S_k})
\ee
where $\beta$ is a 2-homology class on $S_k$ and ${q}_{S_k}=(q_a,q_b,q_1,\ldots,q_k)$ is a formal 
multi-symbol which keeps track of the homology class of the map.  
The integration is performed against the virtual fundamental cycle 
$ [\om_{g,0}(S_k,\beta)]^{vir}$ on the moduli space of stable maps to $S_k$, and 
$R^1\rho_*ev^*K_{S_k}$ is the obstruction bundle on the moduli space defined by the following 
diagram 
\[
\xymatrix{ \om_{g,1}(S_k,\beta) \ar[r]^{{\qquad ev}} \ar[d]_{\rho} & S_k \\
 \om_{g,0}(S_k,\beta) & \\}.
\]
In formula \eqref{eq:GWA} we sum over maps with disconnected domain subject to the 
constraint that no connected component of the domain is mapped to a point. 

Note that there is a {\bf B}-model approach to the computation of topological amplitudes 
\eqref{eq:GWA} based on mirror symmetry and holomorphic anomaly equation 
\cite{KMV,LMW,MNV,MNVW,HST,Hi,Mi} and this problem has also been addressed in \cite{KKV} 
from the point of view of BPS state counting. The {\bf B}-model approach is valid in principle 
for all genera and all degrees but it is quite cumbersome for generic values of the K\"ahler 
parameters. Most computations are performed along a one parameter subspace in the complexified 
K\"ahler moduli space. 

Our goal is to find a closed {\bf A}-model expression for the topological
partition function \eqref{eq:GWA}. 
As opposed to toric del Pezzo surfaces, one cannot directly use localization with respect to a 
torus action because there is no torus action on a generic del Pezzo surface $S_k$.  
However we will show in section two that any such surface $S_k$ is related by complex structure 
deformations to a surface $S_k^0$ which admits a degenerate torus action. 
In principle, since Gromov-Witten 
theory is independent on complex structure deformations, one could try to compute 
the topological partition function at the special point $S_k^0$ in the moduli space of $S_k$. 
This approach is however quite subtle for local Gromov-Witten invariants since $S_k^0$ is 
not a Fano surface, therefore the local Gromov-Witten theory of $S_k^0$ is not well defined. 

In order to effectively implement this idea in practice, we have to consider complex structure 
deformations of a suitable compactification of $X_k$. Then the partition function 
$Z_{X_k}$ can be computed in terms of the residual Gromov-Witten theory of $S_k^0$ plus 
extra corrections at infinity. In section two we will construct two different compactifications 
$Y_k, Z_k$ of $X_k$ and study their complex structure deformations. Exploiting invariance with 
respect to such deformations, we will then find a closed form expression for the partition 
function \eqref{eq:GWA} in section three. 
As a byproduct of this approach we will also find a closed from expression for the residual 
Gromov-Witten theory of certain configurations of $(-2,0)$ curves in a projective threefold. 

In the remaining part of the introduction, we would like to discuss the
physical applications of our results. Local del Pezzo surfaces are usually 
associated to phase transitions in the K\"ahler moduli space of various 
string, M-theory, and F-theory compactifications. More precisely
del Pezzo contractions in Calabi-Yau threefolds are related to 
quantum field theories in four and five dimensions \cite{MS,DKV,KKV:geom}
via geometric engineering. According to \cite{MV} del Pezzo surfaces are also 
related via string duality to small instantons 
transitions in heterotic M-theory. In particular, non-toric del Pezzo surfaces seem 
to be related to exotic physics in four, five and six dimensions 
such as nontrivial fixed points of the renormalization group 
\cite{GMS} without lagrangian description and strongly interacting 
noncritical strings \cite{MV,KMV,LMW,MNV,MNVW}. There is also 
a relation between non-toric del Pezzo surfaces and string junctions 
in F-theory \cite{HI}. In this context, 
certain problems of physical
interest such as counting of BPS states reduce to questions related to 
topological strings on local del Pezzo surfaces. 
The topological string approach is especially useful in the absence of 
a perturbative description of the exotic lower dimensional 
physics. This seems to be usually the case with non-toric del 
Pezzo contractions. 

In order to make this discussion more concrete, note that the topological 
partition function computed in the paper is related via geometric engineering 
to instanton effects and gravitational couplings 
\cite{Ninst,LMN} in four and five dimensional quantum field theories. This relation has
been thoroughly investigated for various toric local Calabi-Yau threefolds 
in \cite{IKPA,IKPB,EK,HIV,AK}, but not for the class of quantum field theories 
engineered by non-toric del Pezzo surfaces. It would be very interesting 
to find a precise connection between our exact results with previous
computations performed in the field theory limit such as \cite{MNV}. 

Another question of physical interest is whether there is a {\bf B}-model 
interpretation for the {\bf A}-model approach adopted in this paper. 
For toric Calabi-Yau manifolds, the topological vertex formalism has been 
given a {\bf B}-model interpretation involving matrix models and 
integrable hierarchies in \cite{ADKMV}. Our exact formulas suggest that 
there should exist a similar interpretation for local-nontoric del Pezzo 
surfaces. In particular it would be very interesting to find a direct 
connection to the {\bf B}-model computations performed in 
\cite{KMV,LMW,MNV,MNVW,HST,Hi,Mi}. This idea is supported by the 
fact that the local mirror geometry of non-toric del Pezzo surfaces 
is governed by a holomorphic curve \cite{MNVW,HI}, just as in the 
toric case. 

Finally, one can also find applications of the formalism developed in this 
paper to black hole entropy and its relations to topological strings 
\cite{OSV}. This connection can lead both to a better understanding of 
the microscopic origins of black hole entropy and to a nonperturbative 
formulation of topological string theory. Exact results along these lines 
have been obtained for local curves in \cite{topbh,AOSV} and local toric surfaces 
in \cite{AJS}. These results are based on a microscopic description of 
black holes in local Calabi-Yau geometries in terms of topological 
Yang-Mills theory. Our results suggest that similar results should 
hold for local non-toric del Pezzo surfaces, although making this 
connection precise may involve certain conceptual puzzles.
Since our method is based on compactifications of the local 
models, one would have to understand how to implement an 
analogous construction in twisted Yang-Mills theory. Moreover, 
in the absence of a torus action on the local surface, it is 
not clear if the microscopic count of black hole microstates 
can be reduced to topological Yang-Mills theory on a fixed 
collection of divisors. One may have to perform an  integration 
over a certain moduli space of divisors, which would give rise 
to various complications. This would make a very interesting 
subject for future work.

\section{Compactification and Complex Structure Deformations} 

Following the strategy outlined in the previous section we will construct two projective 
completions of the Calabi-Yau threefolds $X_k$, $3\leq k\leq 7$ and suitable degenerations 
of their complex structures. 

Let us first discuss the degenerations of the surfaces $S_k$. Note that for $k=3,4$ one can 
degenerate $S_k$ to a toric surface $S_k^0$ obtained by blowing up $3$ and respectively $4$ 
fixed points on $\IF_0$ with respect to a generic torus action. The resulting surfaces can be 
viewed as ruling over $\IP^1$ with two reducible fibers. 
For further reference a reducible fiber consisting of $p$ irreducible components\footnote{Note that 
$C_1,\ldots,C_p$ are the reduced irreducible components of the singular fiber. In principle 
these components may have higher multiplicity. For example a fiber of type 
$(-2,-1,-2)$ is of the form  $f=C_1+2C_2+C_3$ i.e. the  middle component has multiplicity $2$.}
$C_1,\ldots,C_p$, $p\geq 2$, with intersection matrix 
\[ 
\left[
\begin{array}{ccccccc} 
n_1 & 1 & 0 & 0 & \ldots & 0 & 0 \cr 
1 & n_2 & 1 & 0 & \ldots & 0 & 0 \cr 
\vdots & \vdots & \vdots & \vdots & \vdots & \vdots & \vdots \cr
0 & 0 & 0 & 0 & \ldots & 1 & n_p \cr 
\end{array} 
\right]
\]
will be called a fiber of type $(n_1,\ldots,n_p)$. 

For $k=3$ we have a reducible fiber of type $(-2,-1,-2)$ and a reducible 
fiber of type $(-1,-1)$ while 
for $k=4$ we have two identical reducible fibers of type $(-2,-1,-2)$
as shown in fig. 1. 

\begin{figure}[ht]
\psfrag{dP2}[c][c][1][0]{{{\normalsize $S_4^0$}}}
\psfrag{dP1}[c][c][1][0]{{{\normalsize $S_3^0$}}}
\centerline{\epsfxsize=11.1cm \epsfbox{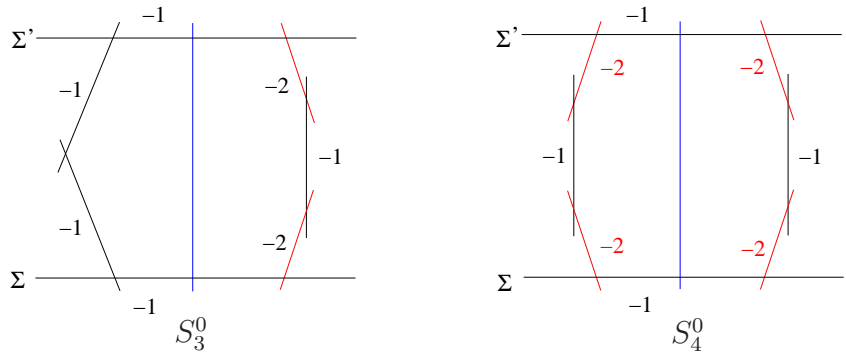}}
\caption{Curve configurations for $dP_4$ and $dP_5$ degenerations.} 
\label{fig:dPcurves} 
\end{figure}

The degenerations for $k=5,6$ can be obtained by exchanging a smooth fiber of $k=4$ with a fiber of 
type $(-1,-1)$ or $(-2,-1,-2)$, respectively; the $k=5,6,7$ degenerations are shown in fig. 2. The 
resulting surfaces $S_k^0$ with $k=5,6,7$ are not toric, but admit a degenerate torus action fixing two 
sections of the ruling pointwise. Note that in principle one can consider many types of 
degenerations of a del Pezzo surface which admit torus actions. However, for reasons that 
will be explained below we are interested in degenerations $S^0_k$
with at most $(-2)$ curves so that the anticanonical bundle $-K_{S^0_k}$ is nef. The self-intersection 
numbers of the canonical sections and fiber components are recorded in fig. 1 and fig. 2. 

\begin{figure}[ht]
\psfrag{dP4}[c][c][1][0]{{{\normalsize $S_6^0$}}}
\psfrag{dP5}[c][c][1][0]{{{\normalsize $S_7^0$}}}
\psfrag{dP3}[c][c][1][0]{{{\normalsize $S_5^0$}}}
\centerline{\epsfxsize=15.5cm \epsfbox{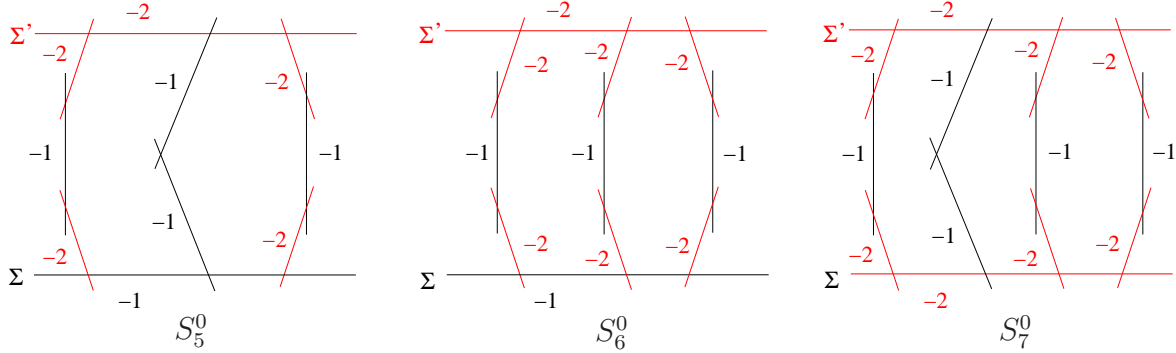}}
\caption{Curve configurations for $dP_6$, $dP_7$ and $dP_8$ degenerations.} 
\label{fig:dP8lego} 
\end{figure}

\noindent Next we discuss the projective completions $Y_k, Z_k$. 

\subsection{Projective Bundle Completion}

A straightforward compactification of $X_k$ is given by the 
projective bundle $Y_k =\IP(K_{S_k}\oplus \CO_{S_k})$. The projection map $Y_k\to S_k$ has
two canonical sections $S_k^+, S_k^-$ with normal bundle $N_{S_k^+/Y_k}\simeq K_{S_k}$ and 
respectively $N_{S_k^-/Y_k}\simeq -K_{S_k}$. $X_k$ is isomorphic to the complement of 
$S_k^-$ in $Y_k$. From now on we will refer to $S_k^-$ as the divisor at infinity on $Y_k$.  
Obviously, $Y_k$ is not Calabi-Yau. In fact a short computation shows that 
\be\label{eq:cancls}
K_{Y_k} = -2S_{k}^-
\ee
is supported on the divisor at infinity. In particular we have \[-K_{Y_k}\big|_{S_k^-} \simeq 
-2K_{S_k}.\] 
Therefore $-K_{Y_k}$ is nef since $-K_{S_k}$ is ample. 

The topological partition function of $Y_k$ is given by 
\be\label{eq:GWB} 
Z_{Y_k} = \sum_{g\in \IZ} \sum_{\stackrel{\beta\in H_2(Y_k)}{\beta\neq 0}}
g_s^{2g-2} q_{Y_k}^{\beta}\int_{[\om_{g,0}(Y_k,\beta)]} 1
\ee
where $\beta$ is a curve class on $Y_k$ and the integration is performed against the 
virtual fundamental class on the moduli space of stable maps of $Y_k$. 
Note that contrary to the common practice in Gromov-Witten theory for non Calabi-Yau 
target spaces, we have not included more general correlators in the definition 
of the partition function. If $Y_k$ were Fano, we would simply have $Z_{Y_k}=1$. 
However in the present case $Y_k$ is not Fano, and we claim that the partition function 
\eqref{eq:GWB} reduces to \eqref{eq:GWA}. 

Indeed, recall that the virtual dimension of the moduli space of stable maps to $Y_k$ without marked 
points is given by 
\[ 
-\langle K_{Y_k}, \beta\rangle 
\] 
where $\langle \ ,\ \rangle $ denotes the pairing between cohomology and homology. 
Any curve class $\beta \in H_2(Y_k)$ is of the form 
\[ 
\beta = nf + \iota^{+}_*(\beta^+)
\]
where $n\in \IZ$, $n\geq 0$, $\iota^+:S_k \to Y_k $ denotes the section of the projective 
bundle $Y_k \to S_k$ mapping $S_k$ to $S_k^+$, and $\beta^+\in H_2(S_k)$ is a curve class 
on $S_k$. Using the formula \eqref{eq:cancls}, it follows that the virtual dimension of the 
moduli space is $2n\geq 0$. 
By construction, only curve classes with virtual dimension zero have a nontrivial contribution to 
the partition function \eqref{eq:GWB}, therefore the claim follows. 

We take the degeneration $Y^0_k$ of $Y_k$ to be the projective bundle 
$\IP(K_{S^0_k}\oplus \CO_{S^0_k})$ over the ruled surface $S^0_k$. Since Gromov-Witten theory is
invariant under complex structure deformations, the partition function \eqref{eq:GWB} must be 
equal 
to the partition function of $Y_k^0$ 
\be\label{eq:GWC} 
Z_{Y^0_k} = \sum_{g\in \IZ} g_s^{2g-2} \sum_{\stackrel{\beta\in H_2(Y_k^0)}{\beta\neq 0}}
q_{Y^0_k}^{\beta}\int_{[\om_{g,0}(Y_k^0,\beta)]} 1
\ee
up to a simple of variables $q_{Y^0_k}=q_{Y^0_k}(q_{Y_k})$ which will be written down 
explicitly in section 
three. This change of formal K\"ahler parameters reflects the change in the Mori cone of $Y_k$ as 
its complex structure degenerates to $Y^0_k$. 

Following the notation introduced above, we have two canonical sections $\iota^{\pm} : S_k^0 
\to Y_k$ with images $(S^0_k)^\pm$. We will also denote by $X^0_k$ the complement of $(S^0_k)^-$ 
in $Y^0_k$. 
Now the main observation is that the partition function of $Y^0_k$ can be evaluated by 
localization with respect to a degenerate torus action. To explain the choice of the torus action, 
note that there are two distinct torus actions on the total space $X^0_k$ of the canonical 
bundle $K_{S^0_k}$ which will be referred to as the diagonal and respectively antidiagonal 
action in the following. 

The diagonal action is induced by the canonical torus action on $S^0_k$ which rotates the fibers 
of the ruling $S^0_k\to \IP^1$ leaving two sections $\Sigma, \Sigma'$ pointwise fixed. 
The antidiagonal action is obtained from the
diagonal action by flipping the sign of the torus weight along the fiber of the canonical class. 
This terminology is in agreement with the conventions of \cite{local}. Let us denote by 
$L_1\oplus L_2$ and respectively $L'_1\oplus L'_2$ the normal bundles to the fixed 
sections $\Sigma, \Sigma'$ in $X^0_k$. We will also denote by $(s_1,s_2)$, 
$(s_1',s_2')$ the weights of the 
induced torus action on $L_1\oplus L_2$ and respectively $L'_1\oplus L'_2$. 
Then the diagonal action on $X^0_k$ induces the diagonal action on 
i.e.
\[ 
s_1-s_2=0, \qquad s_1'-s_2'=0
\]
while the antidiagonal action on $X^0_k$ induces the antidiagonal action
\[
s_1+s_2=0,\qquad s'_1+s'_2=0
\]
on $L_1\oplus L_2$ and respectively $L'_1\oplus L'_2$.

Since $X_k^0$ is the complement of the divisor at infinity in $Y^0_k$, the above torus action will 
extend canonically to $Y^0_k$. In the following we will use the antidiagonal torus action in order
to perform localization computations in Gromov-Witten theory. 
By construction, the canonical bundle of $Y^0_k$ is 
given by 
\be\label{eq:canclsB}
K_{Y^0_k} = -2(S^0_k)^{-}.
\ee
Repeating the previous arguments, we can show again that only curve classes of the form 
$\beta=\iota^+_*(\beta^+)$, $\beta^+\in H_2(S^0_k)$ 
will contribute to the partition function. 
The main difference with respect to the generic case is that now we can have effective 
curves $C\subset (S^0_k)^{+}$ which move in $Y^0_k$ along normal directions to $(S^0_k)^{+}$.
Therefore we can have in principle classes $\beta^-\in H_2(S^0_k)$ so that 
\[\iota^-_*(\beta^-) \subset \iota^+_*(H_2(S^0_k)),\] which were absent in the 
generic case. Note however that curve classes of the form $\beta= nf + \iota_*^+(\beta^+)$, $n>0$
still do not contribute to the partition function since the corresponding moduli spaces 
have positive virtual dimension. 
Therefore the generic fixed loci in the moduli space of stable maps to $Y^0_k$ 
which contribute to the partition function \eqref{eq:GWC} are of the form 
\[
\Xi^+\times \Xi^-
\] 
where $\Xi^\pm$ are fixed loci in the moduli spaces $\om_{g^\pm,0}(S^0_k,\beta^\pm)$.
This is the main reason we allow at most $(-2)$ 
curves on the degenerations $S^0_k$. In the presence of $(-m)$ curves in the base, with $m>2$, 
curve classes will $n>0$ can contribute to the partition function, 
and the fixed loci will have a more complicated structure. 

Then the partition function factorizes into two parts corresponding to the residual
Gromov-Witten theory of the sections $(S^0_k)^{\pm}$
\be\label{eq:GWD}
Z_{Y^0_k} = Z_{(S^0_k)^{+}}Z_{(S^0_k)^{-}}
\ee 
where 
\[ 
\begin{aligned}
& Z_{(S^0_k)^{+}} = \sum_{g\in \IZ} \sum_{\stackrel{\beta^+\in H_2(S^0_k)}{\beta^+\neq 0}}
g_s^{2g-2} q_{Y^0_k}^{\beta^+}\int_{[\om_{g,0}(S^0_k,\beta^+)]_{T^+}} {e_T(R^1\rho_*ev^*K_{S^0_k})\over 
e_T(R^0\rho_*ev^*K_{S^0_k})},\cr
& Z_{(S^0_k)^{-}} = \sum_{g\in \IZ} \sum_{\stackrel{\beta^-\in H_2(S^0_k)\setminus\{0\}}
{\iota^{-}_*(\beta^-)\subset \iota^+_*(H_2(S^0_k))}}
g_s^{2g-2} q_{Y^0_k}^{\beta^-}\int_{[\om_{g,0}(S^0_k,\beta^-)]_{T^-}} {e_T(R^1\rho_*ev^*(-K_{S^0_k}))\over 
e_T(R^0\rho_*ev^*(-K_{S^0_k}))}.\cr
\end{aligned} 
\]
In the above equations, the integration is performed with respect to the equivariant virtual 
fundamental class with respect to the antidiagonal torus action. For further reference it may 
be helpful to emphasize some aspects of these formulas. 

Note that although both canonical sections $(S^0_k)^\pm$ are isomorphic to $S^0_k$, they are 
not equivariantly isomorphic to $S^0_k$, therefore the virtual fundamental classes 
in these two equations are different. More precisely $[\om_{g,0}(S^0_k,\beta^+)]_{T^+}$ denotes the 
equivariant virtual fundamental class with respect to the antidiagonal torus action 
while $[\om_{g,0}(S^0_k,\beta^+)]_{T^-}$ denotes the 
equivariant virtual fundamental class with respect to the diagonal torus action.
Moreover, these two sections have different 
normal bundles in $Y^0_k$ -- namely isomorphic to $K_{S^0_k}$ and respectively $(-K^0_{S_k})$ 
-- which explains the different sign in the construction of the obstruction classes. 
Note also that the summation over homology classes in the second equation is restricted to curve 
classes $\beta^-\in  H_2(S^0_k)$ so that $\iota^{-}_*(\beta^-)$ lies in  
$\iota^+_*(H_2(S^0_k))$. We will show in the next section that such classes must be supported
on configurations of smooth irreducible $(-2)$ rational curves on $S^0_k$. In the following we 
will often refer to $Z_{(S^0_k)^-}$ as corrections at infinity. In section three will compute 
the partition function \eqref{eq:GWD} using the ruled vertex formalism of \cite{DFS}. 
We conclude this section with a construction of a different projective completion of 
$X_k$ which will also play an important role in the next section. 

\subsection{Elliptic Fibration Completion}

A second completion of $X_k^0$ -- which is more natural in physics -- can be obtained by embedding 
$S^0_k$ in a compact Calabi-Yau threefold $Z_k$. We will take $Z_k$ to be a smooth elliptic fibration with 
a section over $S_k$, or, in other words, a smooth Weierstrass model with base $S_k$.
Then $S_k$ is naturally embedded in $Z_k$ by the canonical section 
$\sigma : S_k \to Z_k$. By a slight abuse of notation we will also denote by $S_k$ the image of 
this section in $Z_k$. The distinction should be clear from the context. 

Since $S_k$ is Fano, any effective curve on $S_k$ does not move along the normal directions to $S_k$ in 
$Z_k$. Therefore the partition function \eqref{eq:GWA} represents in fact a subsector of the Gromov-Witten
theory of $Z_k$. More precisely we can rewrite 
\be\label{eq:GWE} 
Z_{X_k} = \sum_{g\in \IZ} \sum_{\stackrel{\beta\in \sigma_*(H_2(S_k))}{\beta\neq 0}} g_s^{2g-2} 
q_{Z_k}^{\beta} \int_{[\om_{g,0}(Z_k,\beta)]}1.
\ee
Next we construct a degeneration of $Z_k$ by deforming the base $S_k$ to the surface $S^0_k$ as in 
subsection 2.1. The smooth Weierstrass model over $S_k$ will deform to a smooth Weierstrass $Z^0_k$
model over $S_k^0$ since $-K_{S_k^0}$ is nef. We will denote by $\sigma:S_k^0\to Z_k^0$ the unique section. 
Exploiting again the invariance of Gromov-Witten theory under complex structure deformations, 
we find that the partition function \eqref{eq:GWE} must be equal to 
\be\label{eq:GWF}
Z_{S^0_k \subset Z^0_k} = \sum_{g\in \IZ} \sum_{\stackrel{\beta\in 
\sigma_*(H_2(S^0_k))}{\beta\neq 0}}g_s^{2g-2}
q_{Z^0_k}^{\beta} \int_{[\om_{g,0}(Z_k^0,\beta)]}1
\ee
up to a change of formal variables $q_{Z^0_k} = q_{Z^0_k}(q_{Z_k})$. 
Note that since $-K_{S^0_k}$ is nef but not Fano, there are curves $C\subset S_k^0$ which move in $Z^0_k$ 
along the normal directions to $S^0_k$. Therefore the partition function \eqref{eq:GWF} receives contributions
from curves in $Z_k^0$ which are not in $S^0_k$. These contributions will be analyzed in the next section. 

\section{Ruled Vertex and Local del Pezzo Surfaces} 

In this section we derive a closed expression for the topological partition function of local 
del Pezzo surfaces using the complex structure degenerations constructed in section two. 
The starting point is formula \eqref{eq:GWD}. The residual partition function $Z_{(S^0_k)^+}$ 
can be computed using the ruled vertex formalism developed in \cite{DFS} since $S^0_k$ is a 
rational ruling equipped with the antidiagonal torus action. Therefore we are left with the 
computation of the corrections at infinity $Z_{(S^0_k)^-}$. Recall that $Z_{(S^0_k)^-}$
represents the residual partition function of the surface $(S^0_k)^-\subset Y^0_k$ 
with respect to the torus action defined above equation \eqref{eq:canclsB}. This partition function 
cannot be directly evaluated by one of the known vertex techniques because the infinitesimal 
neighborhood of $(S^0_k)^-$ in $Y^0_k$ is not $K$-trivial. In principle one could try to develop 
a new vertex technique for this case using the Donaldson-Thomas/Gromov-Witten correspondence 
\cite{MNOP} but this would take us too far afield. Fortunately there is another approach 
to the computation of $Z_{(S^0_k)^-}$ based on a subtle interplay between the two complex structure 
degenerations introduced in section two which reduces the problem to a familiar situation. 
We will show in subsection 3.1  that the 
computation of $Z_{(S^0_k)^-}$ reduces to the computation 
of the residual Gromov-Theory for certain trees of $(-2)$ curves embedded in $(S^0_k)^+$
which can be evaluated using vertex techniques. 

\subsection{Corrections at Infinity } 

By construction, $Z_{(S^0_k)^-}$ is generated by effective curves on 
$Y^0_k$ contained in the divisor at infinity 
$(S^0_k)^-$ whose homology class lie in the image $\iota^+_*(H_2(S^0_k))$. 
Suppose $C^-$ is an irreducible effective curve in $(S^0_K)^-$ and let $C\subset S^0_k$ 
denote the image of its projection to $S^0_k$.  We will also denote by $C^+$ the image of $C$
under the section $\iota^+:S^0_k\to Y^0_k$. 

The restriction of the projective bundle $Y^0_k\to S^0_k$ to $C$ is a projective bundle 
$\IP(K_{S^0_k}\big|_{C} \oplus \CO_C)$ isomorphic to a ruled surface of degree 
\[
-\hbox{deg}(K_{S^0_k}\big|_{C}). 
\]
Since $-K_{S^0_k}$ is nef, $-\hbox{deg}(K_{S^0_k}\big|_{C})\geq 0$. 
This ruled surface has two canonical sections which can be identified with $C^+,C^-$ with self-intersection
numbers 
\[
(C^+)^2 = \hbox{deg}(K_{S^0_k}\big|_{C}),\qquad 
(C^-)^2 = -\hbox{deg}(K_{S^0_k}\big|_{C}).
\]
Let $f$ be the class of the fiber of the ruling. Then we have 
\[ 
C^- = C^+ - \hbox{deg}(K_{S^0_k}\big|_{C}) f ,
\]
and it follows that the class of $C^-$ on $Y^0_k$ is in the image $\iota^+_*(H_2(S^0_k))$ 
if and only if 
\[
\hbox{deg}(K_{S^0_k}\big|_{C}) =0.
\]
This singles out $C$ as an irreducible rational $(-2)$ curve on $S^0_k$. 
Therefore we can conclude that all curves classes contributing to $Z_{(S^0_k)^-}$ must be linear
combination with nonnegative integral coefficients of $(-2)$ curve classes on $S^0_k$. 
There are finitely many such curves on $S^0_k$ hence the corrections at infinity can be identified with the 
residual Gromov-Witten theory of a finite collection of $(-2,0)$ curves on $Y^0_k$ embedded in 
$(S^0_k)^-$. The relevant configurations of curves for the surfaces $S^0_k$, $3\leq k \leq 7$ are represented 
with red line segments in fig. 1 and 2. 

Although $Y^0_k$ is not a Calabi-Yau threefold, all these curves are local Calabi-Yau curves since they have 
normal bundle isomorphic to $\CO_{\IP^1}\oplus \CO_{\IP^1}(-2)$. However the induced torus action on their 
normal bundles is not the anti-diagonal action, therefore their contributions cannot be computed using the 
topological vertex formalism. We will evaluate them 
employing a different strategy. 

Note that the residual partition function of the section $(S^0_k)^+$ factorizes naturally as a product 
of two factors 
\be\label{eq:GWG} 
Z_{(S^0_k)^+} = Z_{(S^0_k)^+}^{>0} Z_{(S^0_k)^+}^{0}
\ee
where 
\be\label{eq:GWH}
\begin{aligned}
& Z_{(S^0_k)^{+}}^{>0} = \hbox{exp}\bigg[\sum_{g=0}^\infty \sum_{\stackrel{\beta^+\in H_2(S^0_k)}
{\langle -K_{S^0_k}, \beta^+\rangle >0 }}
g_s^{2g-2} q_{Y^0_k}^{\beta^+}\int_{[\om_{g,0}(S^0_k,\beta^+)]_{T^+}} {e_T(R^1\rho_*ev^*K_{S^0_k})\over 
e_T(R^0\rho_*ev^*K_{S^0_k})}\bigg],\cr
& Z_{(S^0_k)^{+}}^{0} = \hbox{exp}\bigg[\sum_{g=0}^\infty \sum_{\stackrel{\beta^+\in H_2(S^0_k)}
{\langle -K_{S^0_k}, \beta^+\rangle =0 }}
g_s^{2g-2} q_{Y^0_k}^{\beta^+}\int_{[\om_{g,0}(S^0_k,\beta^+)]_{T^+}} {e_T(R^1\rho_*ev^*K_{S^0_k})\over 
e_T(R^0\rho_*ev^*K_{S^0_k})}\bigg].\cr
\end{aligned} 
\ee
As opposed to the convention used so far, the integration in the exponent is now performed on the 
moduli space of stable maps with connected domain. The first equation in \eqref{eq:GWH} 
represents the contribution of curve classes which have positive intersection product 
with the anticanonical class to the residual partition function of $(S^0_k)^+$. Since these curves cannot 
move in the normal directions to $(S^0_k)^+$ in $Y^0_k$, it follows that the moduli spaces 
$\om_{g,0}(S^0_k,\beta^+)$, with $\langle -K_{S^0_k}, \beta^+\rangle >0$ are in fact compact. Therefore 
the contribution $Z_{(S^0_k)^{+}}^{>0}$ does not depend on the choice of the torus action. 
The second factor $Z_{(S^0_k)^{+}}^{0}$ represents the contributions of curve classes which are orthogonal 
to the anticanonical class. These curves can be deformed in normal directions to $(S^0_k)^+$ in $Y^0_k$,
hence $Z_{(S^0_k)^{+}}^{0}$ depends on the choice of a torus action. 
Similarly, the contribution of curves at infinity $Z_{(S^0_k)^{-}}$ also depends on the choice of torus 
action. Summarizing this paragraph, we conclude that the partition function of a local del Pezzo surface
admits the following factorization 
\be\label{eq:GWI} 
Z_{X_k} = Z_{(S^0_k)^+}^{>0} Z_{(S^0_k)^+}^{0}Z_{(S^0_k)^{-}}.
\ee

Now let us concentrate on the elliptic fibration completion described in section 2.2.
The partition function \eqref{eq:GWF} can be written analogously as a product of two factors 
\be\label{eq:GWJ} 
Z_{S^0_k\subset Z^0_k} = Z_{S^0_k\subset Z^0_k}^{>0}Z_{S^0_k\subset Z^0_k}^0
\ee
where $Z_{S^0_k\subset Z^0_k}^{>0}$ represents the contribution of curve classes which have 
positive intersection with the anticanonical class $K_{S^0_k}$, and $Z_{S^0_k\subset Z^0_k}^0$ 
represents the contribution of curve classes orthogonal to the anticanonical class. 

The main observation is that the following identity holds 
\be\label{eq:identityA} 
Z_{(S^0_k)^+}^{>0}=Z_{S^0_k\subset Z^0_k}^{>0}
\ee 
since both factors have identical intrinsic definitions in terms of the ruled surface $S^0_k$. 
Since by the deformation argument the left hand side of \eqref{eq:GWJ} must be equal to the local del 
Pezzo partition function $Z_{X^0_k}$, we are left with the following identity 
\be\label{eq:identityB}
Z_{(S^0_k)^+}^{0}Z_{(S^0_k)^{-}}= Z_{S^0_k\subset Z^0_k}^0.
\ee
Note that by construction the right hand side of equation \eqref{eq:identityB} depends a priori on the choice 
of a torus action on $Y^0_k$, while the left hand side is intrinsically defined in terms of $Z^0_k$. 
In fact it is not hard to evaluate $Z_{S^0_k\subset Z^0_k}^0$. 

The curve classes which contribute to $Z_{S^0_k\subset Z^0_k}^0$ are of the form $\beta \in H_2(S^0_k)$, 
$\langle K_{S^0_k},\beta\rangle =0$. We have already shown that all such curve classes are generated over 
nonnegative integers by a finite collection of $(-2)$ curves on $S^0_k$. 
Suppose $C\subset S^0_k$ is an irreducible rational $(-2)$ curve on $S^0_k$. Since $C$ is orthogonal to 
the anticanonical class, if the elliptic fibration over $S^0_k$ is generic, $C$ does not intersect the 
discriminant $\Delta \subset S^0_k$ which lies in the linear system $|-12K_{S^0_k}|$. 
Therefore the restriction of the elliptic fibration $Z^0_k\to S^0_k$ to $C$ must be isomorphic to a 
direct product $C\times E$, where $E$ is an elliptic curve. 
This holds for any $(-2)$ curve $C\subset S^0_k$. In conclusion, by restricting the elliptic fibration 
to the configurations of $(-2)$ curves, we obtain similar configurations of surfaces of the form $\IP^1\times E$ 
intersecting along elliptic curves. 

Note that for a given $k$ we have several such connected trees of elliptic ruled surfaces on $Z^0_k$
in one-to-one correspondence with the trees of $(-2)$ curves on $S^0_k$. We will generically denote the collection 
of all such surfaces by $\cup_a F_a$ keeping in mind that the structure of the trees and the range of values of 
$a$ depend on $k$. The Mori cone of each such surface is generated by two curve classes associated to the rational ruling and 
respectively the elliptic curve class of the direct product. 
The factor $Z_{S^0_k\subset Z^0_k}^0$ in \eqref{eq:identityB} represents the contribution of all rational 
fiber classes to the topological partition function of $Z_k^0$. 

We claim that this contribution is trivial 
\be\label{eq:GWK} 
Z_{S^0_k\subset Z^0_k}^0=1.
\ee
This can be shown by analyzing the structure of stable maps to $Z^0_k$ with connected domain whose homology class 
is a linear combination of the rational fiber classes of the elliptic curves. Suppose $f:\Sigma \to Z^0_k$ is 
such a map. Then its image must be isomorphic to a connected tree $C=\cup_a C_a$ of $(-2,0)$ curves on $Y^0_k$
so that each irreducible component $C_a$ of the tree is a fiber in the rational ruling of one of the surfaces 
$F_a$. Let $\beta =\sum_a d_a [C_a]$ denote the homology class of this map, where $d_a\geq 0$ are arbitrary 
degrees. Note that the elliptic curve $E$ acts freely by translations on the moduli space $\om_{g,0}(Z_k^0,\beta)$. 
This action preserves the virtual fundamental class since the deformation theory of such a map $f:\Sigma \to Z^0_k$ 
is invariant under translations along the elliptic fiber. Therefore all Gromov-Witten invariants of nontrivial degree 
$\beta\neq 0$ vanish by analogy with the Gromov-Witten invariants of abelian varieties or direct products of the 
form $E\times K3$. 
Substituting this result in \eqref{eq:identityB}, we obtain 
\be\label{eq:identityC} 
Z_{(S^0_k)^{-}}= (Z_{(S^0_k)^+}^{0})^{-1}.
\ee
The right hand side of equation \eqref{eq:identityC} can be evaluated using vertex techniques. 

In conclusion, the evaluation of the right hand side of equation \eqref{eq:GWD} reduces to computations in 
the residual Gromov-Witten theory of the local ruled surface $(S^0_k)^+$, which is captured by the vertex 
formalism of \cite{DFS}. 

\subsection{Ruled Vertex} 

The ruled vertex is the main building block  in the construction of the residual Gromov-Witten theory 
of a local ruled surface $S\to \Sigma$ over a curve $\Sigma$ of genus $g$ where the ruling is allowed to have 
finitely many reducible fibers. Let $X$ denote the total space of the canonical bundle $K_S$. 
There is a canonical torus action on $S$ which fixes two sections $\Sigma, \Sigma'$ pointwise. 
We lift this action to the anti-diagonal action on $X$ described in the paragraph below equation 
\eqref{eq:GWC} in section two. The essential point is that with this choice, the fixed sections 
$\Sigma$, $\Sigma'$ become equivariantly Calabi-Yau curves on $X$ as in \cite{local}. 

The residual Gromov-Witten theory of $S$ with respect to the anti-diagonal torus action can be computed 
using a TQFT algorithm based on a decomposition of the ruled surface in basic building blocks.
This decomposition is induced by a decomposition of the base $\Sigma$ in pairs of 
pants and caps. The caps are centered at the points in the base representing the locations of the 
reducible fibers. Given such a decomposition of $\Sigma$ we obtain a decomposition of the surface 
$S$ by taking inverse images under the projection map $S\to \Sigma$. To each pair of pants $\Lambda\subset 
\Sigma$ and to each cap $\Delta$ we associate the piece of ruled surface obtained by 
restricting the ruling to $\Lambda$ and respectively $\Delta$. Note that this decomposition is 
compatible with the tours action along the fibers of $S$. 

The formalism of \cite{DFS} associates to each such piece of ruled surface a topological partition function 
labeled by two Young diagrams $R,R'$ associated to the two fixed sections. The pair of pants vertex depends on a 
level $p$ which is a topological invariant of the ruling over the pair of pants. Since all surfaces we are interested 
in can be obtained by successive blow-ups of $\IF_0$ we only need the level zero ruled vertex in the 
following, which is given by the formula 
\be\label{eq:ruledvertex} 
V_{RR'}(y) = \left(\sum_Q W_{RQ} W_{QR'} y^{l(Q)}\right)^{-1} 
\ee
where $y$ is the K\"ahler parameter of the ruling. 

We will also need caps, which are in one to one correspondence with the irreducible fiber types introduced in 
section two. The corresponding partition functions $C^{(n_1,\ldots,n_p)}_{RR'}$ can be computed by localization 
using the topological vertex \cite{topvert}. We have the following expression 
\be\label{eq:caps} 
C^{(n_1,\ldots,n_p)}_{R,R'}(y_1,\ldots,y_p) = \sum_{Q_1,\ldots,Q_p} W_{RQ_1}\prod_{i=1}^{p}
\left(q^{n_i\kappa(Q_i)}(-1)^{n_il(Q_i)}y_i^{l(Q_i)}W_{Q_{i}Q_{i+1}}\right)
\ee
where $y_i$ are the exponentiated K\"ahler parameters of the fiber components satisfying $\prod_{i=1}^p {y_i}=y$, 
and by convention $Q_{p+1}=R'$. As explained in \cite{DFS}, reducible fibers can be obtained by blowing-up 
points on the sections $\Sigma,\Sigma'$. In the gluing algorithm one should insert a factor $q^{-\kappa(R)/2}
(-1)^{l(R)}$ for each blow-up centered at a point on $\Sigma$ and a similar factor $q^{-\kappa(R')/2}
(-1)^{l(R')}$ for each blow-up centered at a point on $\Sigma'$. The decompositions for the $S_6^0$ and $S_8^0$ 
degenerations are presented in fig. 3.

\begin{figure}[ht]
\psfrag{S6}[c][c][1][0]{{{\normalsize $S_6^0$}}}
\psfrag{S8}[c][c][1][0]{{{\normalsize $S_8^0$}}}
\psfrag{V}[c][c][1][0]{{{\footnotesize $V_{RR'}$}}}
\psfrag{C}[c][c][1][0]{{{\footnotesize $C^{(-2,-1,-2)}_{RR'}$}}}
\psfrag{D}[c][c][1][0]{{{\footnotesize $C^{(-1,-1)}_{RR'}$}}}
\psfrag{S}[c][c][1][0]{{{\footnotesize $\Sigma$}}}
\psfrag{Sr}[c][c][1][0]{{{\footnotesize {\red $\Sigma'$}}}}
\centerline{\epsfxsize=12.3cm \epsfbox{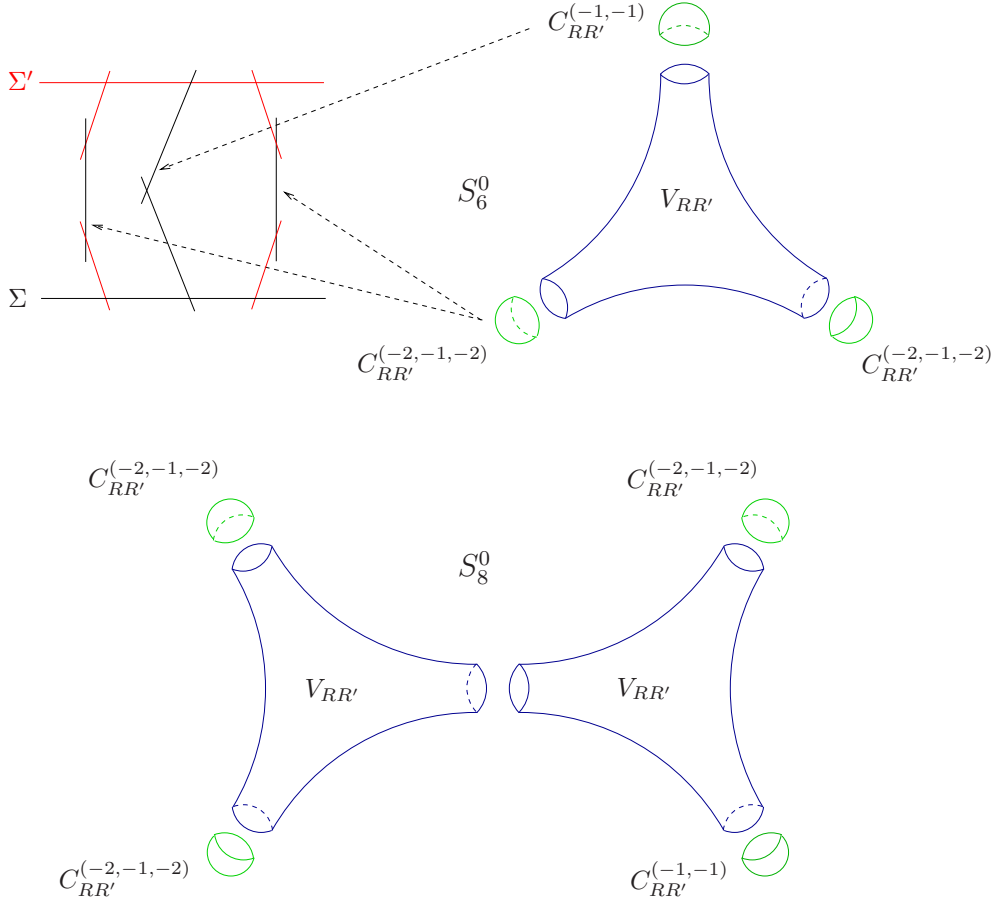}}
\caption{Lego decomposition for $dP_6$ and $dP_8$ degenerations.} 
\end{figure}

Using these building blocks, we can construct the residual partition function of any ruled surface with 
reducible fibers of any type.
One has to multiply all pairs of pants and caps in the decomposition of $S$ 
inserting the formal K\"ahler parameters 
$y_b^{l(R)},(y_b')^{l(R')}$ associated to the fixed sections as well as  the blow-up factors, and sum over $R,R'$. 
The next subsection concludes the computation of the topological partition function for local del Pezzo 
surfaces. 

\subsection{Final Evaluation} 

Let us first evaluate the main contribution $Z_{(S^0_k)^+}$ to the partition function \eqref{eq:GWD}
applying the algorithm described in section 3.2. Note that the surfaces $S_k^0$, $k=3,4$ are toric, therefore 
in these cases the residual partition function can be entirely expressed in terms of the topological vertex. 
For $k\geq 5$, this is not possible and we have to use the ruled vertex. 
We will write the partition function in terms of the K\"ahler parameters $q_a,q_b, q_1,\ldots, q_k$ 
associated to the Mori cone generators of a generic del Pezzo surface $S_k$. As a preliminary step we 
first express the formal K\"ahler parameters $q_{Y^0_k}$ associated to the Mori cone generators of the 
degenerate surface $S^0_k$ in terms of the generic parameters $q_{S_k} = (q_a,q_b,q_1,\ldots,q_k)$, 
$k=3,\ldots,7$. This change of variables is represented graphically in fig. 4. We have:

\be\label{eq:GWM}
\begin{aligned} 
Z_{(S^0_3)^+} = 
&\sum_{R,R'} \big[(-1)^{l(R)+l(R')} q^{-(\kappa(R)+\kappa(R'))/2}q_1^{l(R)} q_3^{l(R')}
V_{RR'}(q_aq_bq_1^{-1}q_3^{-1}) \cr 
& \times C^{(0)}_{RR'}(q_aq_bq_1^{-1}q_3^{-1})
C^{(-1,-1)}_{RR'}(q_aq_1^{-1},q_bq_3^{-1})  C^{(-2,-1,-2)}_{RR'}(q_bq_1^{-1}q_2^{-1},q_2,q_aq_2^{-1}q_3^{-1})\big],  
\cr 
Z_{(S^0_4)^+} = 
& \sum_{R,R'} \big[(-1)^{l(R)+l(R')} q^{-(\kappa(R)+\kappa(R'))/2}q_1^{l(R)} q_3^{l(R')}
V_{RR'}(q_aq_bq_1^{-1}q_3^{-1}) \cr 
& \times C^{(0)}_{RR'}(q_aq_bq_1^{-1}q_3^{-1})
C^{(-2,-1,-2)}_{RR'}(q_aq_1^{-1}q_4^{-1},q_4,q_bq_3^{-1}q_4^{-1})  
C^{(-2,-1,-2)}_{RR'}(q_bq_1^{-1}q_2^{-1},q_2,q_aq_2^{-1}q_3^{-1})\big],\cr
Z_{(S^0_5)^+} = 
& \sum_{R,R'} \big[(-1)^{l(R)} q^{-\kappa(R)/2-\kappa(R')}q_1^{l(R)} (q_3q_5^{-1})^{l(R')}
V_{RR'}(q_aq_bq_1^{-1}q_3^{-1}) \cr 
& \times C^{(-1,-1)}_{RR'}(q_aq_bq_1^{-1}q_3^{-1}q_5^{-1},q_5)
C^{(-2,-1,-2)}_{RR'}(q_aq_1^{-1}q_4^{-1},q_4,q_bq_3^{-1}q_4^{-1})\cr  
& \times C^{(-2,-1,-2)}_{RR'}(q_bq_1^{-1}q_2^{-1},q_2,q_aq_2^{-1}q_3^{-1})\big],\cr
Z_{(S^0_6)^+} = 
& \sum_{R,R'} \big[(-1)^{l(R)} q^{-\kappa(R)/2-\kappa(R')}q_1^{l(R)} (q_3q_5^{-1})^{l(R')}
V_{RR'}(q_aq_bq_1^{-1}q_3^{-1}) \cr 
& \times C^{(-2,-1,-2)}_{RR'}(q_aq_bq_1^{-1}q_3^{-1}q_5^{-1}q_6^{-1},q_6,q_5q_6^{-1})
C^{(-2,-1,-2)}_{RR'}(q_aq_1^{-1}q_4^{-1},q_4,q_bq_3^{-1}q_4^{-1})\cr  
& \times C^{(-2,-1,-2)}_{RR'}(q_bq_1^{-1}q_2^{-1},q_2,q_aq_2^{-1}q_3^{-1})\big],\cr
Z_{(S^0_7)^+} = 
& \sum_{R,R'} \big[ q^{-\kappa(R)-\kappa(R')}(q_1q_7^{-1})^{l(R)} (q_3q_5^{-1})^{l(R')}
(V_{RR'}(q_aq_bq_1^{-1}q_3^{-1}))^2 \cr 
& \times C^{(-2,-1,-2)}_{RR'}(q_aq_bq_1^{-1}q_3^{-1}q_5^{-1}q_6^{-1},q_6,q_5q_6^{-1})
C^{(-2,-1,-2)}_{RR'}(q_aq_1^{-1}q_4^{-1},q_4,q_bq_3^{-1}q_4^{-1})\cr  
& \times C^{(-2,-1,-2)}_{RR'}(q_bq_1^{-1}q_2^{-1},q_2,q_aq_2^{-1}q_3^{-1})
C^{(-1,-1)}(q_7,q_aq_bq_1^{-1}q_3^{-1}q_7^{-1})\big].\cr
\end{aligned}
\ee 

\begin{figure}[!ht]
\begin{center}
\psfrag{a1}[c][c][1][0]{{{\footnotesize $\Sigma'$}}}
\psfrag{a20}[c][c][1][0]{{{\footnotesize $\Sigma$}}}
\psfrag{a19}[c][c][1][0]{{{\footnotesize {\red $\Sigma'$}}}}
\psfrag{a26}[c][c][1][0]{{{\footnotesize {\red $\Sigma$}}}}
\psfrag{a27}[c][c][1][0]{{{\footnotesize $q_1$}}}
\psfrag{a22}[c][c][1][0]{{{\footnotesize $q_2$}}}
\psfrag{a23}[c][c][1][0]{{{\footnotesize $q_4$}}}
\psfrag{a24}[c][c][1][0]{{{\footnotesize $q_5$}}}
\psfrag{a2}[c][c][1][0]{{{\footnotesize $q_3$}}}
\psfrag{a25}[c][c][1][0]{{{\footnotesize $q_6$}}}
\psfrag{a29}[c][c][1][0]{{{\footnotesize $q_7$}}}
\psfrag{a3}[c][c][1][0]{{{\footnotesize $q_bq_3^{-1}$}}} 
\psfrag{a21}[c][c][1][0]{{{\footnotesize $q_aq_1^{-1}$}}} 
\psfrag{a4}[c][c][1][0]{{{\footnotesize{\blue $q_aq_bq_1^{-1}q_3^{-1}$}}}}
\psfrag{a5}[c][c][1][0]{{{\footnotesize{\red $q_aq_2^{-1}q_3^{-1}$}}}}
\psfrag{a6}[c][c][1][0]{{{\footnotesize{\red $q_bq_1^{-1}q_2^{-1}$}}}}
\psfrag{a7}[c][c][1][0]{{{\footnotesize{\red $q_aq_bq_1^{-1}q_3^{-1}q_5^{-1}q_6^{-1}$}}}}
\psfrag{a8}[c][c][1][0]{{{\normalsize $S_4^0$}}}
\psfrag{a9}[c][c][1][0]{{{\normalsize $S_3^0$}}}
\psfrag{a14}[c][c][1][0]{{{\normalsize $S_5^0$}}}
\psfrag{a15}[c][c][1][0]{{{\normalsize $S_6^0$}}}
\psfrag{a16}[c][c][1][0]{{{\normalsize $S_7^0$}}}
\psfrag{a10}[c][c][1][0]{{{\footnotesize{\red $q_bq_3^{-1}q_4^{-1}$}}}}
\psfrag{a11}[c][c][1][0]{{{\footnotesize{\red $q_aq_1^{-1}q_4^{-1}$}}}}
\psfrag{a12}[c][c][1][0]{{{\footnotesize{\red $q_3q_5^{-1}$}}}}
\psfrag{a13}[c][c][1][0]{{{\footnotesize{\black $q_aq_bq_1^{-1}q_3^{-1}q_7^{-1}$}}}}
\psfrag{a17}[c][c][1][0]{{{\footnotesize{\black $q_aq_bq_1^{-1}q_3^{-1}q_5^{-1}$}}}}
\psfrag{a18}[c][c][1][0]{{{\footnotesize{\red $q_5q_6^{-1}$}}}}
\psfrag{a28}[c][c][1][0]{{{\footnotesize{\red $q_1q_7^{-1}$}}}}
\epsfig{file=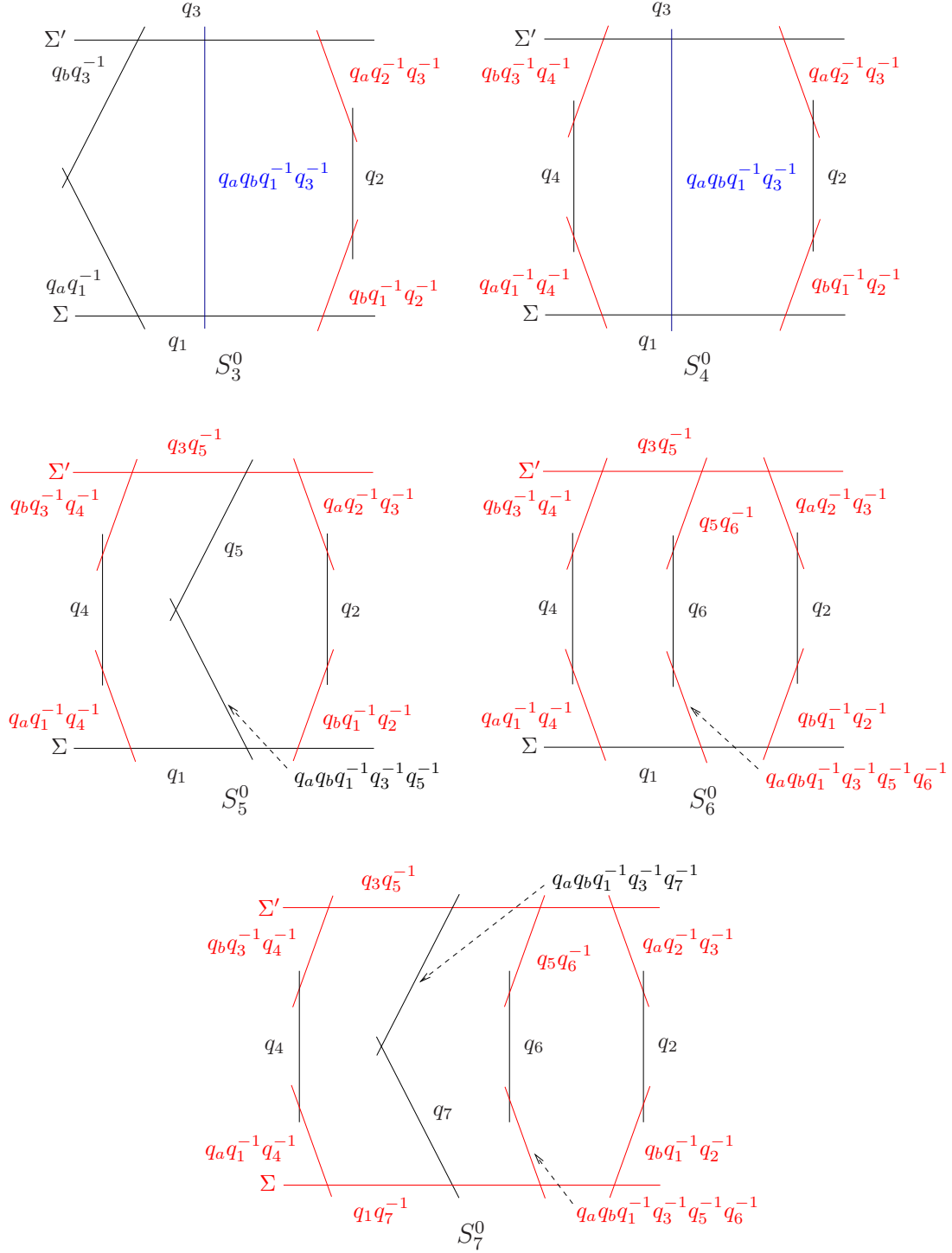,width=5.0in} 
\end{center}
\caption{Mori cone generators for $S_3^0,\ldots, S_7^0$ degenerations.} 
\label{fig:dPmc} 
\end{figure}

In order to finish the computation, we have to calculate the corrections at infinity 
\eqref{eq:identityC}, which are generated by $(-2)$ curves on $S^0_k$. 
The configurations of $(-2)$ curves consist of finite trees 
represented with red line segments in fig. 1 and 2 for all values of $k=3,\ldots,7$.
Note that for $k=3,4$ we encounter only toric trees of $(-2)$ curves which can be evaluated using the 
topological vertex formalism. For $k=5,6,7$ we have nontoric trees as well since at least one component of the 
three intersects three other components. Such configurations are not toric since there is no torus action on 
$\IP^1$ fixing three distinct points. However we can still evaluate the contribution of a nontoric tree 
to the partition function using the degenerate torus action. The contribution of the central component 
can be obtained using the 
level $(0,1)$ pair of pants vertex $P^{(0,1)}$ for local curves constructed in \cite{local}, 
\be\label{eq:localvert} 
P^{(0,1)}=\sum_{R} {1\over W_{R0}}
\ee
Let us record the results for the trees of $(-2)$ curves for all values of $k=3,\ldots,7$. 

\be\label{eq:minustwo}
\begin{aligned}
Z^0_{(S^0_3)^+} = & \sum_{R_1,R_2} (q_bq_1^{-1}q_2^{-1})^{l(R_1)} (q_aq_2^{-1}q_3^{-1})^{l(R_2)} (W_{R_10})^2
(W_{R_20})^2,\cr\nonumber
\end{aligned}
\ee

\be\label{eq:minustwoi}
\begin{aligned}
Z^0_{(S^0_4)^+} = & \sum_{R_1,\ldots R_4} (q_bq_1^{-1}q_2^{-1})^{l(R_1)} (q_aq_2^{-1}q_3^{-1})^{l(R_2)} 
(q_bq_3^{-1}q_4^{-1})^{l(R_3)} (q_aq_4^{-1}q_1^{-1})^{l(R_4)} \cr\nonumber
\end{aligned}
\ee

\be\label{eq:minustwoii}
\begin{aligned}
& \qquad \quad \times (W_{R_10})^2(W_{R_20})^2(W_{R_30})^2(W_{R_40})^2,\cr
Z^0_{(S^0_5)^+} = & \sum_{R_1,\ldots,R_5} (q_bq_1^{-1}q_2^{-1})^{l(R_1)} (q_aq_2^{-1}q_3^{-1})^{l(R_2)} 
(q_bq_3^{-1}q_4^{-1})^{l(R_3)} (q_aq_4^{-1}q_1^{-1})^{l(R_4)} (q_3q_5^{-1})^{l(R_5)}\cr
& \qquad \quad \times (W_{R_10})^2 (W_{R_40})^2 W_{0R_2}W_{R_2R_3}W_{R_3R_5}W_{R_50},\cr
Z^0_{(S^0_6)^+} = & \sum_{R_1,\ldots,R_7} (q_bq_1^{-1}q_2^{-1})^{l(R_1)} (q_aq_2^{-1}q_3^{-1})^{l(R_2)} 
(q_bq_3^{-1}q_4^{-1})^{l(R_3)} (q_aq_4^{-1}q_1^{-1})^{l(R_4)} (q_3q_5^{-1})^{l(R_5)}\cr
& \qquad\quad \times (q_5q_6^{-1})^{l(R_6)} 
(q_aq_bq_1^{-1}q_3^{-1}q_5^{-1}q_6^{-1})^{l(R_7)}(W_{R_10})^2(W_{R_40})^2(W_{0R_7})^2(W_{R_30})^{-1}\cr
& \qquad \quad  \times W_{0R_2}W_{R_2R_3}W_{0R_6}W_{R_6R_3}W_{0R_5}W_{R_5R_3}, 
\cr
Z^0_{(S^0_7)^+} = & \sum_{R_1,\ldots,R_8} 
(q_bq_1^{-1}q_2^{-1})^{l(R_1)} (q_aq_2^{-1}q_3^{-1})^{l(R_2)} 
(q_bq_3^{-1}q_4^{-1})^{l(R_3)} (q_aq_4^{-1}q_1^{-1})^{l(R_4)} (q_3q_5^{-1})^{l(R_5)}\cr
& \qquad\quad \times (q_5q_6^{-1})^{l(R_6)} 
(q_aq_bq_1^{-1}q_3^{-1}q_5^{-1}q_6^{-1})^{l(R_7)}(q_1q_7^{-1})^{l(R_8)}(W_{R_30})^{-1}(W_{R_80})^{-1}\cr
& \qquad \quad \times W_{0R_2}W_{R_2R_3}W_{0R_6}W_{R_6R_3}W_{0R_5}W_{R_5R_3}W_{0R_1}W_{R_1R8}
W_{0R_7}W_{R_7R_8}W_{0R_4}W_{R_4R_8}.\cr
\end{aligned}
\ee

According to equation \eqref{eq:identityC}, the corrections at infinity are obtained by taking the inverse 
of the above formulas. Note that the contribution of an isolated $(-2)$ curve at infinity has been computed 
before by localization in \cite{BP}. Their result is in agreement with our formulas. 
To conclude this section, the topological partition function of a generic nontoric local del Pezzo surface 
is given by the formula 
\be\label{eq:gendP} 
Z_{X_k}= {Z_{(S^0_k)^+}\over Z^0_{(S^0_k)^+}}
\ee 
with $Z_{(S^0_k)^+}, Z^0_{(S^0_k)^+}$, $k=3,\ldots,7$ given in \eqref{eq:GWM}, \eqref{eq:minustwo}. 
We have computed the local Gromov-Witten invariants of $S_5$ using the above formulas for curve classes 
of the form $d_aa+d_bb-\sum_{i=1}^5 d_ie_i$ with $0\leq d_a,d_b\leq 2$ and checked that our results are 
in agreement with the count of BPS states in \cite{KKV}. 

\appendix
\section{Conventions for Quantum Invariants} 

In this section we summarize definitions and conventions for the quantum invariants occurring in 
section 3. 

For a Young diagram $R$ we denote by $l(R)$ the total number of boxes of $R$ and by 
$\kappa(R)$ the following expression 
\[ 
\kappa(R)=  2\sum_{\tableau{1}\in R} (i(\tableau{1})-j(\tableau{1}))
\]
where $i(\tableau{1}),j(\tableau{1})$ specify the position of a given box 
in the Young diagram. 

For two Young diagrams $R,P$, we define $W_{RP}$ to be the formal expression depending on $q=e^{ig_s}$ 
given by 
\[ 
W_{RP}(q) = \hbox{lim}_{N\to \infty} q^{-{N(l(R)+l(P))\over 2}} 
{S_{RP}(q,N)\over S_{00}(q,N)}.
\]
where $S_{RP}(q,N)$ is the modular $S$-matrix acting on the characters of the $U(N)$ affine Lie algebra. 
Note that $W_{PR}$ is symmetric in $(P,R)$ and
can be written  in terms of Schur functions $s_R$:
\[
 W_{P R}(q)=s_{R}\left(q^{-i+1/2} \right)
s_{P}\left(q^{R^i-i+1/2} \right) 
\]
where $R^i$ is the length of the $i$-th row of Young diagram $R$
and $i=1,\ldots, \infty.$
\smallskip

{\it Acknowledgements}. D.-E. D. was partially supported by an Alfred P. Sloan fellowship and the work of B.F. 
was partially supported by DOE grant DE-FG02-96ER40949.

\bibliography{strings,m-theory,susy,largeN}

\providecommand{\href}[2]{#2}\begingroup\raggedright\begin{thebibliography}{10}

\bibitem{topvert}
\newblock M. Aganagic, A. Klemm, M. Mari\~no and C. Vafa, ``The Topological
Vertex,'' Commun. Math. Phys. {\bf 254} 
(2005) 425, {\tt hep-th/0305132}.

\bibitem{ADKMV}
M. Aganagic, R. Dijkgraaf, A. Klemm, M. Marino and C. Vafa, 
``Topological Strings and Integrable Hierarchies,'' 
 Commun.Math.Phys. {\bf 261} (2006) 45, {\tt hep-th/0312085}.

\bibitem{AOSV}
M. Aganagic, H. Ooguri, N. Saulina and C. Vafa, 
``Black Holes, q-Deformed 2d Yang-Mills, and Non-perturbative Topological
Strings,'' Nucl.Phys. {\bf B715} (2005) 304, {\tt hep-th/0411280}. 

\bibitem{AJS}
M. Aganagic, D. Jafferis and N. Saulina, 
``Branes, Black Holes and Topological Strings on Toric Calabi-Yau Manifolds,'' 
{\tt hep-th/0512245}.

\bibitem{BP}
\newblock J. Bryan and R. Pandharipande,
\newblock ``Curves in Calabi-Yau 3-folds and Topological Quantum Field
Theory,'' 
{\tt math.AG/0306316}.

\bibitem{local}
\newblock J. Bryan and R. Pandharipande,
\newblock ``The Local Gromov-Witten Theory of Curves,'' 
{\tt math.AG/0411037}.


\bibitem{AK}
H. Awata and H. Kanno, 
`` Instanton Counting, Macdonald Function and the Moduli Space of D-Branes,''
JHEP 0505 (2005) {\bf 039}, {\tt hep-th/0502061}. 



\bibitem{DFG}
D.-E. Diaconescu, B. Florea and A. Grassi, 
``Geometric Transitions, del Pezzo Surfaces and Open String Instantons,''
Adv.Theor.Math.Phys. {\bf 6} (2003) 643, {\tt hep-th/0206163}. 

\bibitem{DFS}
\newblock D.-E. Diaconescu, B. Florea and N. Saulina, 
\newblock ``A Vertex Formalism for Local Ruled Surfaces,'' 
{\tt hep-th/0505192.}

\bibitem{DF}
\newblock D.-E. Diaconescu and  B. Florea, ``The Ruled Vertex and 
${\tt D}$-${\tt E}$ 
Degenerations,'' {\tt hep-th/0507058.}

\bibitem{DKV} 
 M. R. Douglas, S. Katz and C. Vafa, 
``Small Instantons, del Pezzo Surfaces and Type I' theory,'' 
Nucl.Phys. {\bf B497} (1997) 155, {\tt hep-th/9609071}. 

\bibitem{EK}
T. Eguchi and H. Kanno, 
``Topological Strings and Nekrasov's Formulas,''
JHEP {\bf 0312} (2003) 006, {\tt hep-th/0310235}.

\bibitem{GMS}
O. Ganor, D. R. Morrison and N. Seiberg, 
``Branes, Calabi-Yau Spaces, and Toroidal Compactification of 
the ${\cal N}=1$ Six-Dimensional $E_8$ Theory'', Nucl.Phys. {\bf B487} 
(1997) 93, {\tt  hep-th/9610251}. 


\bibitem{HI}
A. Hanany and A. Iqbal, ``Quiver Theories from D6-branes via Mirror
Symmetry,'' JHEP {\bf 0204} (2002) {\tt hep-th/0108137}. 

\bibitem{THI} 
T. Hauer and A. Iqbal, ``del Pezzo Surfaces and Affine 7-Brane Backgrounds,''
 JHEP {\bf 0001} (2000) 043 {\tt hep-th/9910054}.

\bibitem{HIV} 
T. J. Hollowood, A. Iqbal and C. Vafa, 
``Matrix Models, Geometric Engineering and Elliptic Genera,''
{\tt hep-th/0310272}.  

\bibitem{HST}
\newblock  S. Hosono, M.-H. Saito and A. Takahashi, ``Relative Lefschetz
Action and BPS State Counting,'' 
 Internat. Math. Res. Notices {\bf 15} (2001) 783, {\tt math.AG/0105148}.

\bibitem{Hi}
\newblock S. Hosono, ``Counting BPS States via Holomorphic Anomaly
Equations,'' 
{\tt hep-th/0206206.}

\bibitem{IKPA}
 A. Iqbal and A.-K. Kashani-Poor, `` Instanton Counting and Chern-Simons
 Theory,'' Adv.Theor.Math.Phys. {\bf 7} (2004) 457, {\tt hep-th/0212279}. 

\bibitem{IKPB} 
A. Iqbal and A.-K. Kashani-Poor,
``$SU(N)$ Geometries and Topological String Amplitudes,''
{hep-th/0306032}.

\bibitem{KKV:geom}
S. Katz, A. Klemm and C. Vafa,
``Geometric Engineering of Quantum Field Theories,'' 
 Nucl.Phys. {\bf B497} (1997) 173, {\tt hep-th/9609239}. 

\bibitem{KKV}
\newblock  S. Katz, A. Klemm and C. Vafa, ``M-Theory, Topological Strings and 
Spinning Black Holes,'' Adv. Theor. Math. Phys. {\bf 3} (1999) 1445, {\tt hep-th/9910181}.

\bibitem{KMV}
\newblock  A. Klemm, P. Mayr and C. Vafa, ``BPS States of Exceptional Non-Critical Strings,'' 
{\tt hep-th/9607139}.

\bibitem{LMW}
\newblock  W. Lerche, P. Mayr and N. P. Warner, `` Non-Critical Strings, del
Pezzo Singularities and Seiberg-Witten 
Curves,'' Nucl. Phys. {\bf B499} (1997) 125, 
{\tt hep-th/9612085}.

\bibitem{LMN} 
A. S. Losev, A. Marshakov and N. Nekrasov,
``Small Instantons, Little Strings and Free Fermions,''
{hep-th/0302191}.

\bibitem{MNOP} 
\newblock D. Maulik, N. Nekrasov, A. Okounkov and R. Pandharipande,
\newblock `` Gromov-Witten Theory and Donaldson-Thomas Theory I,'' 
{\tt math.AG/0312059}. 

\bibitem{MNV}
\newblock  J. A. Minahan, D. Nemeschansky and N. P. Warner, ``Investigating
the BPS Spectrum of Non-Critical 
${\tt E}_n$ Strings,'' Nucl. Phys. {\bf B508} (1997) 64, {\tt hep-th/9705237}; 
``Partition Functions for BPS States of 
the Non-Critical ${\tt E}_8$ String,'' Adv. Theor. Math. Phys. {\bf 1} (1998)
167, 
{\tt hep-th/9707149}.
 
\bibitem{MNVW}
\newblock J. A. Minahan, D. Nemeschansky, C. Vafa and N. P. Warner, 
``${\tt E}$-Strings and ${\cal N}=4$ Topological 
Yang-Mills Theories,'' Nucl. Phys. {\bf B527} (1998) 581, {\tt hep-th/9802168}.

\bibitem{Mi}
\newblock K. Mohri, ``Exceptional String: Instanton Expansions and
Seiberg-Witten Curve,'' Rev. Math. Phys. {\bf 14} (2002) 
913, {\tt hep-th/0110121}.

\bibitem{MV} 
\newblock D. R. Morrison and C. Vafa, 
``Compactifications of F-Theory on Calabi-Yau Threefolds--II'', 
 Nucl.Phys. {\bf B476} (1996) 437, {\tt hep-th/9603161.}

\bibitem{MS}
D. R. Morrison and N. Seiberg, 
``Extremal Transitions and Five-Dimensional Supersymmetric Field Theories,''
 Nucl.Phys. {\bf B483} (1997) 229, {\tt hep-th/9609070}.

\bibitem{Ninst}
N. Nekrasov, ``Seiberg-Witten Prepotential From Instanton Counting,''
Adv.Theor.Math.Phys. {\bf 7} (2004) 831, {\tt hep-th/0206161}.

\bibitem{OSV}
H. Ooguri, A. Strominger and C. Vafa, ``Black Hole Attractors and 
the Topological String,''  Phys.Rev. {\bf D70} (2004) 106007, 
{\tt hep-th/0405146}. 

\bibitem{topbh}
C. Vafa, ``Two Dimensional Yang-Mills, Black Holes and Topological Strings,''
{\tt hep-th/0406058}. 



\end{thebibliography}\endgroup
\bibliographystyle{utphys}

\end{document}